\documentclass[sigplan,screen]{acmart}
\copyrightyear{2026}
\acmYear{2026}
\setcopyright{cc}
\setcctype{by}
\acmConference[ASPLOS '26]{Proceedings of the 31st ACM International Conference on Architectural Support for Programming Languages and Operating Systems, Volume 2}{March 22--26, 2026}{Pittsburgh, PA, USA}
\acmBooktitle{Proceedings of the 31st ACM International Conference on Architectural Support for Programming Languages and Operating Systems, Volume 2 (ASPLOS '26), March 22--26, 2026, Pittsburgh, PA, USA}
\acmPrice{}
\acmDOI{10.1145/3779212.3790180}
\acmISBN{979-8-4007-2359-9/2026/03}

\usepackage{xspace}
\usepackage{subcaption}
\usepackage{enumitem}
\newcommand{\name}{{LAER-MoE}\xspace}
\usepackage[linesnumbered,ruled,vlined]{algorithm2e}
\usepackage{listings}
\usepackage{xcolor}

\definecolor{codegray}{rgb}{0.95,0.95,0.95}
\definecolor{codegreen}{rgb}{0,0.6,0}
\definecolor{codepurple}{rgb}{0.58,0,0.82}
\definecolor{backcolour}{rgb}{0.97,0.97,0.97}

\lstdefinestyle{bashstyle}{
    backgroundcolor=\color{backcolour},
    commentstyle=\color{codegreen},
    keywordstyle=\color{blue}\bfseries,
    stringstyle=\color{codepurple},
    basicstyle=\ttfamily\footnotesize,
    breakatwhitespace=false,
    breaklines=true,
    captionpos=b,
    keepspaces=true,
    showspaces=false,
    showstringspaces=false,
    showtabs=false,
    tabsize=2,
    frame=single,
    rulecolor=\color{black!30},
    xleftmargin=2pt,
    xrightmargin=2pt,
    aboveskip=8pt,
    belowskip=8pt
}

\begin{document}

\title{\name: Load-Adaptive Expert Re-layout for Efficient Mixture-of-Experts Training}


\author{Xinyi Liu}
\authornote{School of Computer Science \& Beijing Key Laboratory of Software and Hardware Cooperative Artificial Intelligence Systems, Peking University}
\affiliation{
  \institution{Peking University}
  \city{Beijing}
  \country{China}
}
\email{xy.liu@stu.pku.edu.cn}

\author{Yujie Wang}
\authornotemark[1]
\affiliation{
  \institution{Peking University}
  \city{Beijing}
  \country{China}
}
\email{alfredwang@pku.edu.cn}

\author{Fangcheng Fu}
\authornote{School of Artificial Intelligence, Shanghai Jiao Tong University}
\affiliation{
  \institution{Shanghai Jiao Tong University}
  \city{Shanghai}
  \country{China}
}
\email{ccchengff@sjtu.edu.cn}

\author{Xuefeng Xiao}
\affiliation{
  \institution{Bytedance Seed}
  \city{Beijing}
  \country{China}
}
\email{xiaoxuefeng.ailab@bytedance.com}

\author{Huixia Li}
\affiliation{
  \institution{Bytedance Seed}
  \city{Beijing}
  \country{China}
}
\email{lihuixia@bytedance.com}

\author{Jiashi Li}
\affiliation{
  \institution{Bytedance Seed}
  \city{Shenzhen}
  \country{China}
}
\email{lijiashi@bytedance.com}

\author{Bin Cui}
\authornotemark[1]
\authornote{Institute of Computational Social Science, Peking University (Qingdao)}
\affiliation{
  \institution{Peking University}
  \city{Beijing}
  \country{China}
}
\email{bin.cui@pku.edu.cn}
\renewcommand{\shortauthors}{Xinyi Liu et al.}

\begin{abstract}
Expert parallelism is vital for effectively training Mixture-of-Experts (MoE) models, enabling different devices to host distinct experts, with each device processing different input data. 
However, during expert parallel training, dynamic routing results in significant load imbalance among experts: a handful of overloaded experts hinder overall iteration, emerging as a training bottleneck.

In this paper, we introduce LAER-MoE, an efficient MoE training framework. The core of LAER-MoE is a novel parallel paradigm, Fully Sharded Expert Parallel (FSEP), which fully partitions each expert parameter by the number of devices and restores partial experts at expert granularity through All-to-All communication during training. This allows for flexible re-layout of expert parameters during training to enhance load balancing. In particular, we perform fine-grained scheduling of communication operations to minimize communication overhead. Additionally, we develop a load balancing planner to formulate re-layout strategies of experts and routing schemes for tokens during training. We perform experiments on an A100 cluster, and the results indicate that our system achieves up to 1.69x acceleration compared to the current state-of-the-art training systems. 
Source code available at \url{https://github.com/PKU-DAIR/Hetu-Galvatron/tree/laer-moe}.
\end{abstract}

\ccsdesc[500]{Computing methodologies~Artificial intelligence}
\ccsdesc[500]{Computing methodologies~Distributed computing methodologies}

\keywords{Mixture-of-Experts, Load Balancing, Distributed Training, Expert Parallelism, Large Language Models}


\maketitle

\section{Introduction}
\label{sec:intro}

Large Language Models (LLMs) have demonstrated remarkable capabilities across various tasks
~\cite{transformer, survey-llm, gpt, llama2, llama3, qwen3, llm-based-agents-for-tool-learning, db-gpt, deep-learning-for-code-generation, openba}. 
However, the exponential growth in model parameters has led to significant computational and memory challenges in training these models~\cite{ai-memory-wall}. 
Mixture of Experts (MoE) architectures have emerged as a promising solution, enabling the training of models with trillions of parameters while maintaining computational demands by activating only a subset of experts for each input token~\cite{moe, moe-survey}. 
Specifically, the feed-forward network (FFN) layers are typically replaced with MoE layers, which consist of a gating network and multiple FFNs, each representing a different expert. 
In an MoE layer, each token is routed by the gating network to only a few selected experts. The final output is computed as a weighted sum of the outputs from these selected experts.

Despite the computational advantages, the tremendous size of expert parameters often exceeds the memory capacity of individual devices. Therefore, expert parallelism~\cite{gshard,tutel} is a widely used technique to facilitate distributed training of MoE models. In expert parallelism, each device stores only a subset of experts to reduce memory consumption, while other model parameters are replicated across all devices, with different training data allocated to each device.
In each MoE layer, tokens are routed by the gating network to the devices hosting the chosen experts. The outputs of experts are then sent back to the original device of each token.

\begin{figure}[t]
    \centering
    \includegraphics[width=0.47\textwidth]{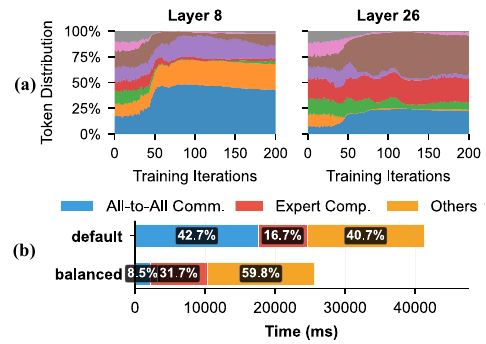}
    \caption{\small{(a) Token distribution during training Mixtral 8x7B, showing significant imbalance. (b) Time breakdown, where ``default'' denotes the profiling result without auxiliary loss, 
    and ``balanced'' denotes the result when enforcing fully balanced routing. }}
    \label{fig:token_time_distribution}
\end{figure}

Though expert parallelism alleviates the memory burden of a single device, it still faces critical challenges when employed for distributed training.
The \textbf{dynamic and imbalanced routing distribution} during training creates significant load imbalances across devices, leading to substantial tail latency that severely impacts overall training efficiency~\cite{fastermoe, smartmoe, flexmoe}. 
For instance, Fig.~\ref{fig:token_time_distribution}(a) records the changes in routing distribution over time while training Mixtral 8x7B, 
showing that overloaded experts emerge almost at every training iteration, and the expert load dynamically fluctuates across iterations. 
Consequently, this imbalance leads to significant tail latency, as overloaded experts take longer to compute, causing other devices to wait. 
This is reflected in notable communication delays, as shown in Fig.~\ref{fig:token_time_distribution}(b). The imbalance in load causes the proportion of All-to-All communication time to increase from less than 10\% to over 40\%.

A notable line of efforts to address this challenge is to enforce load balance through algorithmic techniques like adding auxiliary losses~\cite{gshard, switch-transformers}. 
However, as we will empirically show in Sec.~\ref{sec:algorithm_level} and Sec.~\ref{sec:end-to-end}, algorithmic solutions substantially slow down training convergence~\cite{pangu_ultra}. Thus, this work concentrates on the system-level perspective.

Existing system-level load balancing methods mainly focus on how to adjust the \textit{expert layout}, which can be generally categorized into expert replication and expert relocation.
In vanilla EP (e.g., GShard~\cite{gshard}), expert placement is fixed throughout training.
Expert replication techniques, such as FasterMoE~\cite{fastermoe} and Prophet~\cite{prophet}, replicate high-load experts across multiple devices to mitigate imbalance. 
Expert relocation strategies, like SmartMoE~\cite{smartmoe}, dynamically adjust expert placement based on historical routing distributions, while FlexMoE~\cite{flexmoe} combines both approaches by dynamically adjusting expert replicas and their locations. 
However, these methods suffer from fundamental limitations.

Particularly, these optimizations introduce \textit{additional re-layout overhead, which refrains existing methods from achieving load balance}. 
Expert replication requires additional communication for gradient synchronization of replicated experts. This communication is imbalanced because experts are not replicated across all devices. 
Expert relocation involves migrating expert parameters and optimizer states, which requires a communication volume typically $6\times$ the size of the expert parameters. 
This substantial overhead causes larger peak memory due to the simultaneous allocation of send and receive buffers.
Prior works treat re-layout as an individual phase, forcing a trade-off between load-balancing effectiveness and re-layout overhead.
To manage such overhead, existing methods have to restrict themselves when adjusting the replication and/or relocation strategies.
For instance, SmartMoE regulates relocation frequency to be low (e.g., hundreds of iterations), while Prophet and FlexMoE penalize adjustment strategies that will incur high overhead. 
As a result, existing methods fail to achieve timely, optimal adjustments in response to changes in routing distribution.

To fill this gap, our work explores optimizing expert re-layout strategies in response to dynamically changing routing distributions, while minimizing the impact from expert re-layout to achieve superior load balancing strategies. 
Given the dynamic nature of routing distributions, it is imperative for expert re-layout algorithms to swiftly adapt and adjust strategies. 
Moreover, when expert replication is involved in location strategies, it is essential to efficiently determine which replica each token should be routed to, necessitating rapid expert routing decisions. 
Our primary challenge lies in effectively managing the additional communication and devising enhanced load balancing algorithms, alongside designing efficient expert re-layout algorithms and corresponding routing strategies.

To tackle the challenges previously outlined, our work proposes \name, a novel MoE training system leveraging \textit{\textbf{L}oad-\textbf{A}daptive \textbf{E}xpert \textbf{R}e-layout}.
In contrast to prior works that treat re-layout as an isolated phase, \name innovatively fuses re-layout with FSDP's parameter pre-fetching and gradient reduce-scattering. 
This design enables re-layout at every iteration without additional memory cost and perfectly hides re-layout overhead, providing more flexible re-layout strategies and more immediate load-balancing responses to routing distribution changes.
Specifically, we propose two key innovations to achieve this.

\begin{figure*}[!t]
    \centering
    \begin{minipage}{0.25\linewidth}
      \centering
      \includegraphics[width=\linewidth]{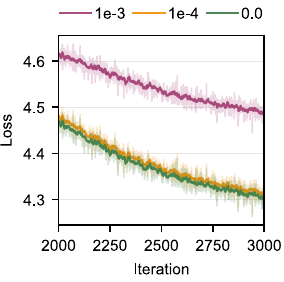}
      \caption{\small Loss curve with different auxiliary loss weights.}
      \label{fig:loss_with_aux}
    \end{minipage}
    \hfill
    \begin{minipage}{0.7\linewidth}
      \centering
      \includegraphics[width=\linewidth]{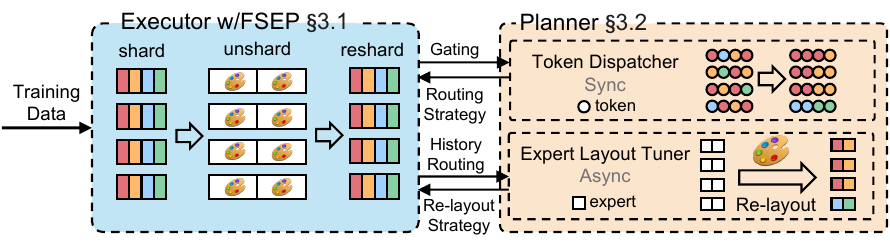}
      \caption{\small The overview of \name.}
      \label{fig:architecture}
    \end{minipage}
\end{figure*}

First, we combine the principles of fully sharded data parallelism (FSDP)~\cite{pytorch-fsdp} with expert parallelism (EP) to introduce a novel parallel paradigm termed Fully Sharded Expert Parallelism (FSEP). 
Unlike traditional EP, which allocates different experts to separate devices, FSEP fully shards each expert, with each device storing only a shard of all experts. 
During forward and backward propagation, each device gathers complete parameters of required experts.
Moreover, FSEP fuses the expert re-layout process with the parameter pre-fetching and gradient reduce-scattering in FSDP, so that the re-layout overhead can be overlapped by the computation time of forward and backward passes.
This offers much better flexibility for designing load balancing algorithms.

Second, we design an intelligent load balancing planner that dynamically monitors expert load during training and generates load-adaptive expert re-layout strategies and token routing results for each training iteration. Our planner employs heuristic greedy strategies to execute an asynchronous expert layout tuner that dynamically resolves expert layout strategies, coupled with a synchronous token dispatcher that rapidly determines expert routing strategies based on real-time training demands, achieving superior load balancing, free from the constraints imposed by re-layout overhead.

Based on the proposed new MoE parallel paradigm FSEP and the intelligent load balancing planner, we efficiently implement our system \name atop PyTorch.
Experiments demonstrate that \name achieves up to 1.69$\times$ acceleration over state-of-the-art (SOTA) training systems. \name's ability to dynamically adjust expert layout during training without communication overhead makes it particularly effective for large-scale MoE model training.

The key contributions of this work include:

\begin{itemize} [leftmargin=*]
    \item We propose a novel MoE parallel paradigm FSEP, enabling flexible, in-training expert re-layout while masking re-layout communication overhead.
    
    \item We devise an intelligent load balancing planner that dynamically co-optimizes token routing and expert layout, achieving timely and effective adjustment.

    \item We develop an efficient MoE training system, namely \name, delivering up to 1.69$\times$ end-to-end acceleration over existing SOTA systems.
\end{itemize}
\section{Preliminary}

\textbf{Mixture of Experts.}
The core idea of Mixture of Experts (MoE) is to sparsely route input data to a subset of expert networks and combine their outputs through a gating mechanism~\cite{moe}.
Specifically, given an input $x$, the output of an MoE model can be expressed as: $y = \sum_{i=1}^{n} g(x)_i \cdot f_i(x),$ where $g(x)_i$ is the output of the gating network representing the weight assigned to the $i$-th expert, $f_i(x)$ is the output of the $i$-th expert network, and $n$ is the total number of experts.
The gating function $g(x)_i$ is often implemented via a top-k selection followed by a softmax operation, i.e., $g(x) = \text{Softmax}(\text{TopK}(x \cdot W_g)),$ where $W_g$ is the weight of the gating network.
This gating mechanism ensures that only a small number of experts (typically k=1 or k=2) are activated for each input.

\textbf{Parallelism in Distributed MoE Training.}
MoE models typically contain far more parameters than dense LLMs, making distributed parallelism essential for efficient training.

\textit{Data parallelism (DP)}~\cite{pytorch-ddp, ai-computing-systems-for-large-language-models-training} splits data samples and replicates the entire model across devices, involving gradient synchronization during training.
To avoid redundant parameter storage, Sharded Data Parallelism (SDP; e.g., DeepSpeed-ZeRO~\cite{zero}, PyTorch FSDP~\cite{pytorch-fsdp}) partitions parameters across devices while preserving DP semantics, enabling training larger models at the cost of extra communication for parameter gathering and gradient scattering.
To overlap these communications with computation, PyTorch FSDP manages communication operations on separate CUDA streams. 
During forward passes, it prefetches parameters of the next layer while computing the current layer. Similarly, during backward passes, it synchronizes gradients of the previous layer while computing gradients for the current layer.

\textit{Expert parallelism (EP)}~\cite{gshard} is designed for MoE models, where different experts are distributed across multiple devices, and each device hosts a subset of experts.
Meanwhile, other non-expert modules are replicated and executed in DP-style.
In each MoE layer, tokens are routed to devices containing the selected experts based on the routing results of the gating network. 
The output from the experts is then returned to the original device of the corresponding token. 
This process involves two communication operations: All-to-All dispatch and All-to-All combine.

\textit{Model parallelism (MP)} splits the model parameters across multiple devices, which can be categorized into tensor parallelism (TP)~\cite{megatron-tp} and pipeline parallelism (PP)~\cite{pipedream, pipedream-flush, pipeline-model-parallelism}. 
TP splits the matrix multiplication computations across devices, utilizing communication to synchronize the computation outputs. 
PP partitions the model into stages with consecutive layers; each device computes its stage and forwards activations to the next via point-to-point communication.

\textit{Heterogeneous Expert Parallelism (HEP)}~\cite{megascale-moe,megatron-moe} is designed to address the load difference between the Attention and MoE layers in MoE models. 
The central idea is to employ distinct parallel strategies for these two types of layers, given their heterogeneous computational and communication characteristics. 
Specifically, the Attention layer typically performs dense computations, making it more suitable for TP.
In contrast, the inherent sparse activation characteristic of MoE layers makes it more adaptable for EP.
\label{sec:algorithm_level}

\textbf{Algorithmic Load Balancing Methods.}
To mitigate the imbalance issue, efforts have been made from the algorithmic perspective.
Involving auxiliary losses is a typical approach to encourage balanced routing~\cite{gshard,switch-transformers}, yet it inevitably impacts model convergence. Fig.~\ref{fig:loss_with_aux} illustrates the convergence curves on Mixtral-8x7b using different auxiliary loss weights. It can be found that the increase in the auxiliary loss weight leads to a rise in the number of steps needed for the model to achieve equivalent performance. 
Some methods further consider discarding tokens to limit the load on each expert~\cite{gshard, tutel}, or directly modifying routing strategies for better balance~\cite{base-layers, base-layers-sinkhorn, hash-layers, expert-choice-routing, bagualu}.
However, these algorithmic approaches represent a trade-off between system efficiency and model quality, and often necessitate extensive experimentation for validation to adjust hyperparameters~\cite{megablocks, pangu_ultra}. 
Our work aims at tackling the imbalance issue from the system-level perspective.

\textbf{System-level Load Balancing Methods.}
To maintain the training unaltered while balancing the load of different devices, existing works have proposed to adjust the expert layout over the devices in response to the dynamic routing distribution~\cite{fastermoe,prophet,smartmoe,flexmoe}. 
Particularly, expert replication and expert relocation are two major techniques, where the former allows high-load experts to be hosted on multiple devices, while the latter dynamically adjusts the locations of experts. 
However, as introduced in Sec.~\ref{sec:intro}, existing works suffer from the overhead caused by expert re-layout and compromise load balance.

\section{System Design}

Fig.~\ref{fig:architecture} illustrates the main architecture of \name, featuring an executor based on a novel parallel paradigm and a load-balancing planner. 
The planner dynamically monitors expert load during training and provides the executor with optimized expert re-layout strategies and specific routing results. 
The executor then employs the new parallel paradigm to implement the updated expert re-layout strategy and routing results in real-time during training.

Specifically, we will first introduce the new parallel paradigm, Fully Sharded Expert Parallelism (FSEP), detailing its execution process, communication volume, and memory cost. 
For the planner, we outline our optimization objectives and propose a heuristic greedy strategy to execute an asynchronous expert layout tuner and a synchronous token dispatcher based on the real-time demands of training. For clarity, the frequently used notations are listed in Tab.~\ref{tab:notation}.

\begin{table}[t]
\centering
\caption{\small{Notations and Definitions used in this paper.}}
\label{tab:notation}
\small
\begin{tabular}{ll}
\toprule
$N$ & Number of devices \\
$E$ & Number of experts \\
$C$ & Expert capacity per device \\
$P$ & Parallel dimensions \\
$\Psi$ & Parameter sizes \\
$R_{i,j}$ & Number of tokens on device $i$ routed to expert $j$ \\
$A_{i,j}$ & Expert re-layout strategy \\
$S_{i,j,k}$ & Token routing strategy \\
$T$ & Time of the cost model \\
$V_{comm/comp}$ & Communication/computation volume per token \\
$B_{inter/intra}$ & Inter-node/intra-node bandwidth \\
$B_{comp}$ &  GPU computational power \\
$F_{ckpt}$ & Checkpoint optimization flag \\
$\text{bw}(i,j)$ & Bandwidth between devices $i$ and $j$ \\
$\text{node}(i)$ & Node where device $i$ is located \\
\bottomrule
\end{tabular}
\end{table}

\subsection{FSEP: Fully Sharded Expert Parallelism}
\label{fsep}
We first introduce the executor, which incorporates the new parallel paradigm FSEP, inspired by EP and FSDP, allowing more flexible retrieval of expert parameters during training. 

\textbf{Workflow of FSEP.}
The fundamental concept of FSEP involves dividing experts into chunks with equal size, corresponding to the number of devices, with each device retaining a distinct chunk from each expert. 
During training, devices restore entire expert parameters through All-to-All communication. 
When restoring parameters, each device does not need to acquire the complete parameters for all experts. Instead, similar to EP, each device restores full parameters from different experts. 

\begin{figure}[t]
\centering
    \includegraphics[width=0.47\textwidth]{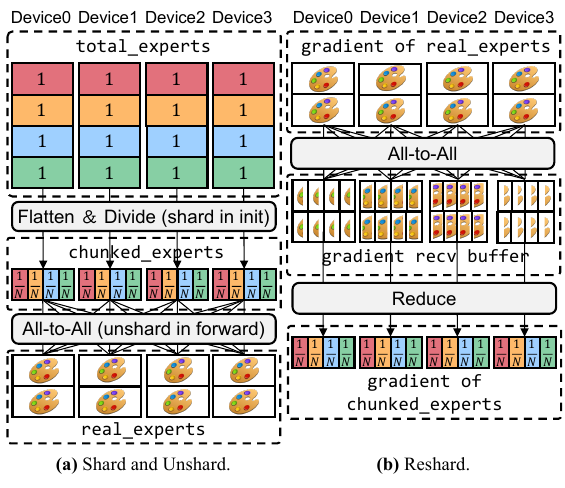}
\caption{\small{Illustration of FSEP, where $N=4$, $E=4$, $C=2$.}}
\label{fig:fsep_sharding}
\end{figure}

Formally, for $N$ devices and $E$ experts, each device has an expert capacity $C$ (the number of complete expert parameters restored), each device stores $E$ chunks, each of size $\frac{1}{N}$ experts, and restores $C$ complete expert parameters during forward and backward. 
Fig.~\ref{fig:fsep_sharding} illustrates such an example, where $N=4$, $E=4$, $C=2$. 
This paradigm mirrors the memory efficiency and initialization approach of FSDP, while during training, it exhibits parallelism similar to EP. 
Thus, we designate this new parallel paradigm as fully sharded expert parallelism (FSEP). 
Furthermore, due to the uniform nature of the communication during restoration, each device can acquire an \textbf{arbitrary layout} of experts, significantly facilitating the application of load balancing methods, which will be discussed in Sec.~\ref{sec:load_balancing_planner}.

To enhance the distributed training of MoE models and mitigate bottlenecks associated with new parallel strategies, FSEP must address a key challenge: 
\textit{How to seamlessly integrate FSEP with existing parallel paradigms, enabling practitioners to easily adopt and transition to the new strategy?}

To tackle this challenge, we draw inspiration from methods in FSDP and refine the existing FSDP codebase. 
To implement unique partitioning logic of FSEP, we introduce three additional sharding methods: \textbf{shard}, \textbf{unshard}, and \textbf{reshard}, as depicted in Fig.~\ref{fig:fsep_sharding}.

\textit{Shard} is invoked during model initialization. 
Similar to traditional FSDP, we treat all expert parameters in the MoE layer as a single unit, and the shard operation is conducted at this level. 
During model initialization, the parameters of each expert are flattened and concatenated into a large parameter, which is then evenly divided into $N$ chunks, with each device retaining a distinct chunk. 
These $E$ partitioned parameters are subsequently concatenated for further management.

\textit{Unshard} is utilized during forward or backward passes to restore complete parameters of the needed experts. 
Since it is unnecessary to acquire all expert parameter information, unshard employs the All-to-All communication operator to obtain the parameters of specified experts.

\textit{Reshard} is primarily used to revert parameters to their partitioned form. 
Since unshard does not release the sharded parameters, no additional operations are required here. 
However, reshard involves re-partitioning and synchronizing complete expert gradients post-backward computation, employing the All-to-All communication operator for gradient synchronization. 
Specifically, each expert gradient is divided into $N$ chunks, which are then sent to corresponding devices for reduction operations, as shown in Fig.~\ref{fig:fsep_sharding}(b).

It is crucial to note that, in traditional FSDP, shard stores the shape of each parameter before flattening as meta information. During unshard, parameters are restored to their unflattened form using split and view operations based on this meta-information, ensuring the correct functioning of the autograd engine of PyTorch. 
In contrast, FSEP requires flattening all expert parameters during shard, yet only restores $C$ expert parameters during unshard, leading to mismatched meta-information. 
To address this, we employ an architecture separating flattened parameters from meta-information, as illustrated in Fig.~\ref{fig:fsep_sharding}(a). 
During shard, we flatten the parameters in \texttt{total\_experts} but record the meta-information of \texttt{real\_experts}, allowing parameters to be accurately restored during unshard, and enabling direct computation using methods in \texttt{real\_experts} during forward computation.

Moreover, it is important to emphasize that FSEP maintains numerical precision identical to FSDP, with no loss in accuracy. This is because FSEP only modifies the parameter storage and communication patterns during sharding and unsharding operations, while the actual forward and backward computations remain unchanged.

\begin{figure}[t]
    \centering
    \includegraphics[width=0.47\textwidth]{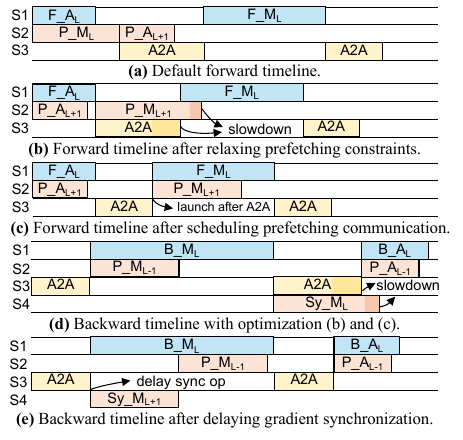}
    \caption{\small{Communication optimization in FSEP, where A represents Attention layer, M represents MoE layer, S represents Stream, blue blocks on S1 represent forward and backward computation (F and B), yellow blocks on S3 represent All-to-All communication of token dispatcher (A2A), and red blocks on S2 and S4 represent prefetching communication (P) and gradient synchronization (Sy).}}
    \label{fig:fsep_communication}
\end{figure}

\textbf{Communication Optimizations.} 
Employing FSEP introduces significant communication overhead. To mitigate this, we implement fine-grained scheduling to obscure this overhead, as shown in Fig.~\ref{fig:fsep_communication}. 
Specifically, the incurred overhead encompasses three operations: All-to-All communication to restore complete expert parameters during unsharding in both forward and backward passes, and gradients resharing and synchronization post-backward computation.

Regarding the communication introduced by unshard, similar to FSDP, we incorporate parameter prefetching in the executor. Contrasting traditional methods that prefetch the parameters of the next unit, within the context of heterogeneous parallelism, this involves prefetching expert parameters during attention computation, resulting in suboptimal overlap effects, as shown in Fig.~\ref{fig:fsep_communication}(a). To address this, we relax the prefetching constraints, overlapping the expert computation of the current layer with the prefetching of expert parameters for the next layer, allowing the more computationally demanding expert computation to obscure the prefetching communication, as depicted in Fig.~\ref{fig:fsep_communication}(b). We further schedule the prefetching communication alongside the All-to-All communication in the token dispatcher, ensuring the former is launched after the latter has concluded, thus reducing potential channel contention, as shown in Fig.~\ref{fig:fsep_communication}(c).

For the communication introduced during resharing, traditional methods rely on the autograd engine to automatically schedule gradient communication, leading to uncontrollable communication timing and overlap effects. To address this, we design a gradient communication delay method, postponing this communication operation when the autograd engine is prepared, deferring it to the subsequent expert layer's backward computation to ensure optimal overlap effects, as illustrated in Fig.~\ref{fig:fsep_communication}(e).

\textbf{Communication and Memory Analysis.}
To demonstrate the effectiveness of the executor, we provide a quantitative analysis of communication volume and memory usage.

\textit{Communication Analysis.}
In FSEP, the communications for shard and reshard are inverse, hence the communication volumes for these operations are consistent. 
Taking shard as an example, the parallel dimension of FSEP is $P_{fsep} = N$, and the size of each expert parameter is $\Psi_{expert}$. 
In each operation, each device needs to send $C$ chunks of $\frac{1}{P_{fsep}}$ of expert parameters to other devices, while receiving $C$ chunks it requires from each of them, leading to a communication volume of $V_{fsep} = C \times (P_{fsep} - 1) \times \frac{1}{P_{fsep}} \times \Psi_{expert}$ per device. The send/receive volumes between any pair of devices are identical. This constitutes a regular, balanced All-to-All communication operation, which achieves the same speed as standard All-to-All communication without incurring additional overhead. 
For the traditional FSDP+EP parallel paradigm, suppose the parallel sizes for EP and FSDP are $P_{ep}$ and $P_{fsdp}$ respectively, when model states and activation values are similar, these dimensions satisfy $P_{ep} \times P_{fsdp} = N$, $C \times P_{ep} = E$. 
During unshard operation of FSDP, each device uses Allgather to aggregate $C$ complete expert parameters across devices within the same FSDP communication group, resulting in a communication volume of $V_{fsdp} = \frac{P_{fsdp} - 1}{P_{fsdp}} \times C \times \Psi_{expert}$.
The ratio of communication volumes between the two methods is $\frac{V_{fsep}}{V_{fsdp}} = \frac{(P_{fsep} - 1) \times P_{fsdp}}{P_{fsep} \times (P_{fsdp} - 1)}$. 
As the cluster size increases, this ratio approaches 1. For instance, when $P_{fsep} = 32$, $P_{ep} = 4$, $P_{fsdp} = 8$, $\frac{V_{fsep}}{V_{fsdp}} \approx 1.1$. 
Compared to traditional paradigms, FSEP incurs only trivial additional communication overhead. Furthermore, through the fine-grained communication scheduling outlined in Fig.~\ref{fig:fsep_communication}, this communication can be overlapped with computation, enabling FSEP to provide dynamic expert re-layout opportunities while retaining overhead similar to the traditional paradigm.

\textit{Memory Analysis.}
To understand the impact of FSEP on model state, we analyze a classic scenario: the MoE layer employs FSEP, whereas other modules utilize equivalently sized FSDP. 
The model state primarily consists of optimizer state, parameter state, and gradient state.
For the optimizer state, since these parallel paradigms are fully sharded, the optimizer state is $\frac{1}{P_{fsep}}$ of the full optimizer state. 
Taking the total parameter size of the complete model as $\Psi_{all}$, and the size of parameters excluding expert parameters for each Transformer layer as $\Psi_{other}$, following the communication optimization in Fig.~\ref{fig:fsep_communication}, all parameters for the next layer are prefetched during the expert computation of current layer, leading to a parameter state overhead of $\frac{\Psi_{all}}{P_{fsep}} + \Psi_{other} + 2 \times C \times \Psi_{expert}$. 
Furthermore, we delayed gradient updates, so the gradient state overhead is consistent with the parameter state. 
Compared to traditional FSDP, our method incurs only an additional $2 \times C \times \Psi_{expert}$ in memory overhead, arising from parameter and gradient states. 
Relative to the full model, this additional memory is negligible.
As for other types of parallel paradigms, they often fail to achieve fully sharded parameter and gradient states, making FSEP a memory-efficient parallel paradigm.
Notably, this additional memory overhead is attributed to the communication optimizations detailed in the previous section, not to the FSEP mechanism itself. Traditional FSDP can similarly incorporate these aforementioned communication optimizations to enhance training efficiency, in which case both FSDP and FSEP exhibit identical memory footprints.

\textbf{Computation-Communication Overlap Analysis.}
To validate the feasibility of overlapping parameter prefetching with expert computation, we provide a theoretical analysis of an MoE layer employing a SwiGLU MLP with bfloat16 precision. Let $H$ denote the hidden dimension, $H'$ the intermediate dimension, $K$ the top-$k$ selection in the router, and $S$ the total token count per device in each micro-batch.

Under the assumption of balanced loading, the computation workload per device is $S \cdot K \cdot (6 \cdot H \cdot H')$, where the parenthesized term indicates the per-token FLOPs for SwiGLU. The communication volume for prefetching (sending and receiving, respectively) by each device is $3 \cdot C \cdot H \cdot H' \cdot \text{sizeof}(\text{bfloat16})$. Consequently, computation latency masks communication overhead provided that the following condition holds:
\begin{equation}
    \small
    \label{eq:computation-communication-overlap}
    S > ({C \cdot V_{comp}})/({K \cdot V_{comm}})
\end{equation}

In our experimental setup (detailed in Sec.~\ref{experiment_setup}), the condition in Eq.~\ref{eq:computation-communication-overlap} is theoretically satisfied when $S \geq 17K$. Empirically, we observe that $S = 16K$ suffices to achieve effective overlap, as the practical computation time often exceeds the theoretical lower bound due to load imbalance. Such a threshold is readily attainable in production scenarios~\cite{megatron-moe, llama-moe}.

\subsection{Load Balancing Planner}
\label{sec:load_balancing_planner}

This section introduces the load balancing planner associated with the FSEP parallel paradigm, encompassing optimization opportunities presented by the new parallel paradigm, cost modeling, and a heuristic efficient load balancing solution method proposed based on training characteristics.

\begin{figure}[t]
    \centering
    \includegraphics[width=0.47\textwidth]{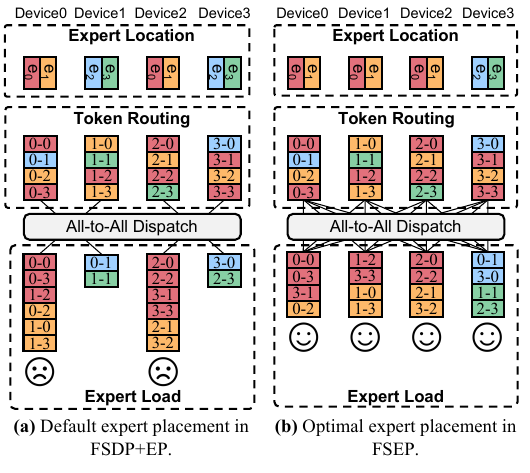}
    \caption{\small{The opportunities of load balancing in FSEP.}}
    \label{fig:fsep_load_balancing}
\end{figure}

\textbf{Optimization Opportunities.}
The major advantage offered by FSEP is the ability to freely select which experts are restored on each device during unshard. 
This allows for the strategic planning of expert re-layout strategies for each iteration, including the number of replicas for each expert and the device where each expert should be restored.
Fig.~\ref{fig:fsep_load_balancing} illustrates such an example: in a traditional setup with $N=4$ and $P_{fsdp} = P_{ep} = 2$, experts 0 and 1 are fixed to be restored on devices 0 and 2, while experts 2 and 3 are fixed on devices 1 and 3. 
However, the actual routing distribution is significantly skewed, with tokens for experts 0 and 1 significantly outnumbering those for other experts, leading to overloading on devices 0 and 2 and underloading on devices 1 and 3. 
With FSEP, it is possible to flexibly adjust the experts restored on device 1 to be experts 0 and 1, subsequently routing part of the tokens for experts 0 and 1 to device 1 for computation, and shifting tokens originally sent to device 1 to device 3, thereby achieving load balancing. 
Thus, FSEP can dynamically alter expert re-layout strategies based on real-time load, promoting load balance across devices, reducing tail latency, and ultimately accelerating training speed.

\textbf{Problem Formulation.}
In the MoE training process, the execution time of the MoE layer is determined by the All-to-All communication for routing distribution and expert computation. Therefore, when modeling the load balancing problem, it is essential to consider both computation and communication costs.
Formally, given $R_{i,j} \in \mathbb{N}^{N,E}$, where $R_{i,j}$ represents the number of tokens on device $i$ routed to expert $j$, expert capacity limit $C$, inter-node communication bandwidth $B_{inter}$, intra-node communication bandwidth $B_{intra}$, communication volume per token $V_{comm}$, computational power of each GPU $B_{comp}$, and computation volume per token $V_{comp}$. 
We define the expert re-layout strategy $A_{i,j} \in \{0,1\}^{N,E}$ and token routing strategy $S_{i,j,k} \in \mathbb{N}^{N,E,N}$, where $A_{i,j}=1$ indicates expert $j$ is placed on device $i$, and $A_{i,j}=0$ otherwise. 
$S_{i,j,k}$ represents the number of tokens on device $i$ routed to expert $j$ that need to be sent to device $k$. 
The optimization goal is to minimize the total time, encompassing both communication and computation time.

To model the communication time, we need to calculate the total cost of All-to-All communication. In particular, we consider the point-to-point communication cost for each pair of devices based on the topology and sums them up, i.e., $T^{comm} = 4 \times V_{comm} \sum_{i,j,k}\frac { S_{i,j,k}} {\text{bw}(i,k)}$, where ${\text{bw}(i,k)}$ represents the intra- or inter-node bandwidth between devices $i$ and $k$, and the multiplier is 4 since the forward and backward passes of an MoE layer involve four All-to-All communication operations in total.

To model the computation cost, which involves the forward and backward computation time, it is common practice to treat the backward cost as twice the forward one. 
Meanwhile, due to the imbalance issue, we need to find the maximum computation time across all devices. 
Thus, we model the computation cost as $T^{comp} = (3 + F_{ckpt}) \times \max_i{T_i^{fw\_comp}}$, where $F_{ckpt} \in \{0,1\}$ indicates whether activation checkpointing (a.k.a. recomputation) is used, and $T_i^{fw\_comp}$ denotes the forward computation time on device $i$. 
Specifically, it is calculated as $T_{i}^{fw\_comp} = V_{comp} \times \sum_{j,k}\frac { S_{k,j,i}} {B_{comp}}$.

Based on these, we propose a joint optimization problem:

{
\small
\begin{align}
\label{eq:optimization_problem}
&\arg\min_{A_{i,j} \in \{0,1\}^{N,E}, S_{i,j,k} \in \mathbb{N}^{N,E,N}} T = T^{comm} + T^{comp} \\
\text{s.t.} &\sum_{i} A_{i,j} = C \quad \forall j \in [0, E-1] \label{eq:constraint_1} \\
&\sum_{k} S_{i,j,k} \times A_{k,j} = R_{i,j} \quad \forall i,j \in [0,N-1] \times [0,E-1] \label{eq:constraint_2}
\end{align}
}

The constraint \ref{eq:constraint_1} limits the number of experts placed on each device, while the constraint \ref{eq:constraint_2} guarantees that all tokens are routed to the correct experts.

\textbf{Solving Algorithm}
The formulated joint optimization problem is a nonlinear integer programming (NIP) problem, which is hard to solve within polynomial time and typically requires existing solvers such as SCIP~\cite{scip} to find approximate solutions. 
Moreover, the problem involves $N^2 \times E + N \times E$ variables, and as the problem size expands, the solver's time consumption grows exponentially. 
In traditional EP, tokens are simply routed to the device hosting the corresponding expert. 
However, FSEP permits expert replication, necessitating additional decisions regarding which replication of the corresponding expert each token should be routed to.
Additionally, there is a strict sequential dependency between the token routing strategy and All-to-All dispatch communication, indicating that token communication dispatch can only proceed once the specific token routing strategy is established. Therefore, it is crucial to derive the specific routing strategy as quickly as possible.
Consequently, directly solving this joint optimization problem is unsuitable. 
To tackle this, we propose a method that separates expert re-layout strategies from token routing strategies. We begin by crafting a token dispatcher that employs a lite routing algorithm based on the immediate dependency of routing strategies. 
Subsequently, we design a heuristic greedy algorithm to solve the expert re-layout strategy based on the lite routing, serving as our expert layout tuner.

\begin{algorithm}[t]
    \caption{\small{Expert Relocation Algorithm}}
    \label{algo:expert_relocation}
    \small
    \LinesNumbered
    \KwIn{$expert\_rep$, $expert\_loads$, $N$, $E$, $C$}
    \KwOut{$A$}
    $expert\_count, device\_loads \leftarrow \text{zeros}(N), \text{zeros}(N)$\;
    $A, list \leftarrow \text{zeros}(E, N), []$\;
    \For{$i \in [0, E-1]$}{
        $list.extend([(i, \frac{expert\_loads[i]}{expert\_rep[i]})] \times expert\_rep[i])$\; \label{code:get}
    } 
    $list \leftarrow DescendingSortByLoad(list)$\; \label{code:sort}
    \For{each $(expert, load)$ in $list$}{
        $node\_cnt \leftarrow \text{count replicas of}~expert~\text{per node}$\;
        $min\_node \leftarrow \min(node\_cnt)$\;
        $available \leftarrow \{i : expert\_count[i] < C \land node(i) \in min\_node\}$\;
        $device \leftarrow \text{select from}~available~\text{with min load}$\;
        $A_{expert,device} \leftarrow A_{expert,device} + 1$\;
        $device\_loads_{device} \leftarrow device\_loads_{device} + load$\;
        $expert\_count_{device} \leftarrow expert\_count_{device} + 1$\;
    }
    \Return $A$\;
\end{algorithm}

\textbf{Token Dispatcher.}
As experts may be replicated, the primary issue in token routing is determining which replica of the expert each token should be routed to.
Our algorithm is inspired by the following two considerations: 
1) In modern training clusters, intra-node communication speeds frequently surpass inter-node communication speeds. Therefore, the routing policy should minimize inter-node token transfers.
2) Fine-grained control over the total number of tokens received by each device requires coordination among all devices, resulting in additional communication overhead, which is inefficient.
Thus, we design a greedy-based lite routing algorithm, which is topology-aware and requires no global routing information, only global expert layout information. 
Specifically, for each expert, if replicas exist within the node, tokens routed to the expert are evenly distributed among all replicas within the node. If no replicas exist within the node, tokens are evenly distributed among all replicas across global nodes.
We detail the lite routing algorithm in Appendix~\ref{sec:append_routing}.

\begin{algorithm}[t]
    \caption{\small{Expert Layout Algorithm}}
    \label{algo:expert_layout}
    \small
    \LinesNumbered
    \KwIn{$R$, $N$, $E$, $C$, $\epsilon$: The size of replicas\_set}
    \KwOut{$A$}
    $replicas\_set \leftarrow \emptyset$\;
    $replicas\_set.append(replica\_allocation(R, N, E, C))$\;
    $evenly\_replicas \leftarrow [\frac{N \times C}{E}] \times E$\;
    $replicas\_set.append(evenly\_replicas)$\;
    \While{$len(replicas\_set) < \epsilon$}{
        $rand\_rep \leftarrow \text{randomly select from } replicas\_set$\;
        $replicas\_set.append( \text{random perturb of } rand\_rep)$\;
    }
    $expert\_load \leftarrow R.sum(axis=0)$\;
    $best\_A, best\_S, best\_T \leftarrow None, None, +\infty$\;
    \For{each $replicas$ in $replicas\_set$}{
        $A \leftarrow expert\_relocation(replicas, R, N, E, C)$\;
        $S \leftarrow lite\_routing(R, A, N, E, C)$\;
        $T \leftarrow \text{time\_cost(A, S)}$\;
        \If{$T < best\_T$}{
            $best\_A, best\_S, best\_T \leftarrow A, S, T$\;
        }
    }
    \Return $best\_A$\;
\end{algorithm}

\textbf{Expert layout tuner.} We now introduce how to solve the expert re-layout strategies using a heuristic greedy approach. Specifically, the problem is divided into two steps: first determining the number of replicas for each expert, and then solving the location of these replicas.

For determining the number of replicas, an intuitive approach is to allocate replicas based on the expert's load: experts with higher loads receive more replicas, while those with lower loads receive fewer. 
Accordingly, we propose a priority-queue-based replica allocation method. 
It detects the expert with the highest average load (i.e., the ratio of actual load to the number of replicas) and expands the number of replicas for this expert until the total number of replicas meets the requirement.
We illustrate the detailed expert replica allocation algorithm in Appendix~\ref{sec:append_replica_alloc}.

For solving expert relocation, we develop a greedy relocation method which is topology-aware and designed with the lite routing strategy to ensure that expert replicas are balanced across different nodes, as detailed in Alg.~\ref{algo:expert_relocation}. 
First, we calculate the average number of tokens each replica of an expert needs to process, then sort all these replicas in descending order (Line~3-5).
Next, we sequentially select devices for each replica.
Since tokens are evenly distributed across all replicas within a node during routing, it is crucial to ensure that the number of replicas of the same expert within each node is as balanced as possible. 
Therefore, for each replica, we first select the set of nodes with the fewest replicas of the current expert (Line~7-9), and then choose the device with the least total load within these nodes as the location of this replica (Line~10-13).

Additionally, since this replica number determination method may not guarantee optimality, we introduce a new scheme: assigning the same number of replicas to all experts based on the total number of replicas (Line~3). 
On top of these two allocation schemes, we introduce randomized perturbations to construct a set of replica schemes (Line~5-7).
We then perform location solving for each scheme in the set sequentially, selecting the optimal strategy as the final solution (Line~9-15). 
The final solving process is depicted in Alg.~\ref{algo:expert_layout}.

\subsection{Overall Workflow}

\begin{figure}[t]
    \centering
    \includegraphics[width=0.47\textwidth]{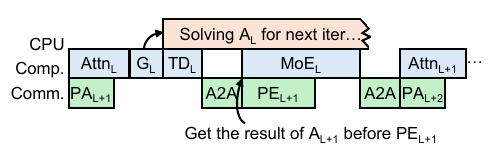}
    \caption{\small{The overall workflow of \name.}}
    \label{fig:overall_workflow}
\end{figure}

Fig.~\ref{fig:overall_workflow} illustrates the overall workflow of \name, where $Attn_L$, $MLP_L$, $G_L$ and $TD_L$ represent the attention layer, MLP in MoE layer, gate layer and token dispatcher of the $L$-th Transformer layer, respectively. $PA_{L+1}$ and $PE_{L+1}$ represent the prefetching of attention parameters and expert parameters for the $L+1$-th Transformer layer.
As solving the expert re-layout strategy does not require high immediacy and requires certain time consumption, we delegate this process to the CPU side. 
During a training process, when computing the current MoE layer, the system first retrieves routing information of the current layer, then dispatches this routing information along with historical data from previous iterations to the expert layout tuner on the CPU side. The CPU references these data to generate new expert re-layout strategies for the next iteration of the layer. 
Meanwhile, the token dispatcher is employed on the GPU side for rapid token routing to facilitate subsequent All-to-All dispatch operations. 
After the All-to-All dispatch operation, the re-layout strategies for the next MoE layer's experts are first retrieved from the CPU, and then the parameters for the next MoE layer are prefetched according to these strategies. 
This prefetch occurs concurrently with MLP computation.
This mechanism continues iteratively layer by layer, organically integrating the executor with FSEP and the load balancing planner.

\section{Implementation}

\name, inspired by Megatron's implementation, extends support for the MoE architecture upon Galvatron~\cite{galvatron_vldb, galvatron_tkde, galvatron_2}, which accommodates various hybrid parallel strategies, including FSDP.
\name expands this foundation with approximately 11k lines of additional Python, CUDA, and C++ code. 
Furthermore, we employ the following methods to further optimize the training process.

\textbf{Heterogeneous Parallel Strategy and Fine-Grained Recomputation.}
Our system also introduces heterogeneous parallel strategies. 
Specifically, for a Transformer layer, FSEP treats the complete expert parameters as a unit, while other parameters in the same layer are placed in another unit. This allows for independent parallel strategy selection, offering more flexible training options.
Additionally, we introduce fine-grained recomputation, which can be performed at the granularity of Attention layers and MLP layers.
For the MoE layer, we also provide the option to recompute only the expert computation part, preventing extra All-to-All communication overhead during recomputation.

\textbf{Customized All-to-All Kernel.}
During expert parameter prefetching and gradient synchronization, 
the inflexibility of torch's all-to-all communication interface results in additional memory for send/receive buffers and extra memory rearrangement overhead.
To address this, we develop a custom communication kernel using CUDA to perform All-to-All communication directly between sharded and unsharded parameters, thereby reducing additional overhead.
Crucially, this custom kernel is designed specifically to avoid memory overhead rather than to improve communication efficiency. It has identical efficiency to standard APIs under the same communication volume.

\textbf{Host Bound Optimization.}
During MoE training, numerous CPU-blocking operations are encountered, such as the \text{masked\_select} operation during traditional token rearrangement and host-to-device and device-to-host operations involved in interactions with the CPU-side solver. 
These operations can block the CPU, delaying subsequent GPU kernel launch instructions and impacting GPU utilization. 
To address this, we modify all host-to-device and device-to-host operations to be asynchronous and orchestrate synchronization timing to ensure correctness and efficiency.
Specifically, we employ dedicated CUDA streams for H2D and D2H transfers, and explicit synchronization is performed on the respective streams only when the data is strictly required. 
This design ensures that the overhead from these operations is negligible.
Additionally, we develop a triton kernel for token rearrangement, avoiding the use of torch operations that can cause blocking. 

\textbf{Efficient Solver of the Planner.}
In the planner, token routing demands immediacy, and the solving of expert re-layout strategies must be executed rapidly to ensure scalability. 
Consequently, we develop an efficient Triton kernel for the token routing algorithm, ensuring rapid execution on the GPU. 
For solving expert re-layout strategies, we create an efficient C++ core and employ a multiprocess approach, utilizing a separate process to interact with the executor.

\section{Experiments}
\label{experiment}

\subsection{Experimental Setups}
\label{experiment_setup}

\textbf{Baseline Systems.}
We compare our system with Megatron~\cite{megatron-moe}, the state-of-the-art distributed training framework supporting MoE models. Megatron supports heterogeneous expert parallelism for flexible strategy combinations. Additionally, for a comprehensive comparison, we extended EP based on Torch FSDP as another baseline training framework with a fully sharded model state.
Specifically, this configuration applies both FSDP and EP to MoE layers while utilizing FSDP for non-MoE layers, as employing standalone EP on MoE layers results in Out-of-Memory (OOM) errors due to insufficient parameter sharding.
Furthermore, we incorporate the communication optimizations detailed in Sec.~\ref{fsep} into this FSDP+EP baseline, thereby isolating and highlighting the efficacy of our approach in addressing load imbalance.
We also compare our approach with FlexMoE~\cite{flexmoe}, which is currently the state-of-the-art load balancing strategy employing expert replication.

\textbf{Hardware Environments.}
We conduct our experiments on a 4-node GPU cluster, with each node containing 8 NVIDIA A100 80GB GPUs. Within nodes, GPUs within a node are connected via NVLink, and nodes are interconnected via Infiniband. The peak unidirectional communication bandwidth intra-node is 300 GB/s, and inter-node is 800 Gbps.

\begin{figure*}[t]
    \centering
    \includegraphics[width=0.99\textwidth]{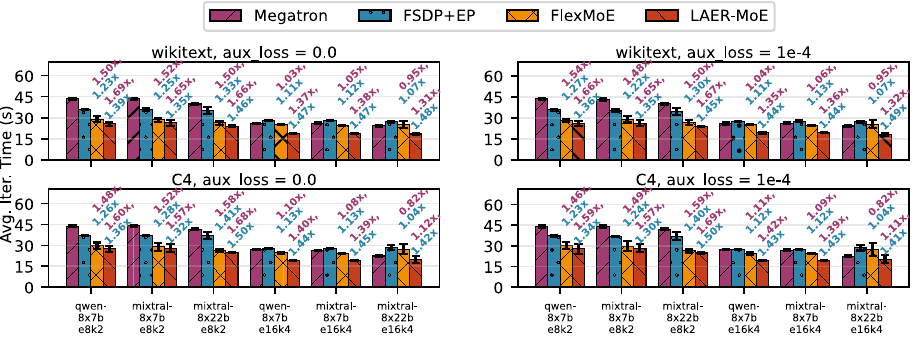}
    \caption{   
    \small{
    End-to-end performance under various model architectures for specific auxiliary loss weights and datasets, as depicted in the subfigures. Speedup ratios compared to Megatron (blue, top) and FSDP+EP (purple, bottom). 
    }
    }
    \label{fig:end_to_end_performance}
\end{figure*}

\begin{table}[t]
    \centering
    \caption{\small{Configurations of the evaluated models.}}
    \small
    \label{tab:model_info}
    \begin{tabular}{ccccc}
        \toprule
        Model & Layers & Params & Activs & E\&K \\
        \midrule
        Mixtral-8x7B & 32 & 46.70B & 12.88B & 8\&2 \\
        Mixtral-8x22B & 18 & 45.46B & 12.86B & 8\&2 \\
        Qwen-8x7B & 32 & 46.69B & 12.88B & 8\&2 \\
        Mixtral-8x7B & 24 & 35.09B & 9.73B & 16\&4 \\
        Mixtral-8x22B & 14 & 35.46B & 10.09B & 16\&4 \\
        Qwen-8x7B & 24 & 35.09B & 9.73B & 16\&4 \\
        \bottomrule
    \end{tabular}
\end{table}

\textbf{Experimental Workloads.}
We experiment on three different model architectures: Mixtral-8x7B, Mixtral-8x22B~\cite{mixtral}, and additionally, transform Mixtral-8x7B into the Qwen-8x7B model architecture~\cite{qwen2.5}. 
Besides the standard configuration of 8 experts with topk being 2(e8k2), we expand these models into configurations of 16 experts with topk being 4(e16k4), without altering the parameter count and computational load per layer. The specific model architectures, total parameter counts, and activated parameter counts are detailed in Tab.~\ref{tab:model_info}, where E\&K denotes the number of experts and topk. 
We utilize the complete Mixtral-8x7B and Qwen-8x7B model for the e8k2 series models, while Mixtral-8x22B has some layers reduced due to memory constraints. 
For the e16k4 series models, as their activations occupy more memory during training, we partially reduce some layers of all three model architectures.
We set the expert capacity per device $C$ to 2 for the e8k2 series models, and 4 for the e16k4 series models.
We conduct our training using a dropless approach and employ Wikitext and C4 as our datasets.
In our experiments, the key hyperparameters influencing efficiency are global batch size, sequence length, and auxiliary loss weight. 
Other hyperparameters (e.g., learning rate, weight decay) do not fundamentally alter the efficiency.
Accordingly, we specify sequence length and auxiliary loss weight (default is 0) in each experiment, while maintaining the standard LLM training configuration for other parameters.

\subsection{End-to-End Performance}
\label{sec:end-to-end}
We evaluate all systems at 8K context across all models and datasets: 20-step warmup, then the average of the next 50 iterations. 
We also compare the impact of auxiliary loss on our method, contrasting no auxiliary loss with using SwitchTransformer auxiliary loss, with a weight of 1e-4. 
Megatron and FSDP+EP are manually tuned to their optimal parallel strategies. 
For FlexMoE (no open-source release), we reproduce its scheduler and replace our expert re-layout planner, comparing it in conjunction with FSEP. 
As shown in Fig.~\ref{fig:end_to_end_performance}, in all scenarios, \name consistently outperformed other methods, achieving up to 1.69$\times$ acceleration over Megatron and 1.50$\times$ over FSDP+EP.
Compared to FlexMoE, LAER-MoE achieves a speedup of up to 1.39$\times$, with an average improvement of 1.20$\times$.

We first compare \name with SOTA systems. 
As shown in Fig.~\ref{fig:end_to_end_performance}, FSDP+EP performs better on e8k2 series models, while Megatron excels with e16k4 series models. 
For e8k2 series models, the larger parameter size results in higher model states memory, requiring Megatron to use a larger TP in attention layer to meet memory limits, hurting efficiency; 
FSDP+EP benefits from fully sharding, using the saved model states memory to increase tokens per micro-batch, thus outperforming Megatron.
For e16k4 series models with slightly fewer parameters, Megatron is allowed to use a smaller TP for improved efficiency, surpassing FSDP+EP.
Both baselines, however, suffer from expert-load imbalance issue, which inflates MoE tail latency, hurting overall efficiency.
\name addresses this by dynamically adjusting the expert layout during training, balancing token computation across devices and alleviating tail latency, achieving optimal results.

Next, we compare \name with FlexMoE. 
FlexMoE attains strong speedups on e8k2 models but weaker gains on e16k4.
This is because e8k2 models have fewer experts, allowing FlexMoE's iterative search to find a relatively optimal expert layout quickly, whereas the larger expert space on e16k4 models limits the quality of its solutions.
Besides, FlexMoE considers the extra adjustment cost and penalizes layout changes, thereby excluding potentially optimal solutions.
In contrast, \name directly adjusts the expert layout each step, offering greater flexibility and consistently achieving better results. 
Note that our FSEP paradigm is decoupled from load balancing solvers, allowing compatibility with any form of planner without imposing additional constraints, enabling the planner to provide more flexible expert layouts for better results. 
We also aim to develop more efficient and effective planners as part of our future work.

\begin{figure}[t]
    \centering
    \begin{subfigure}[b]{0.47\textwidth}
        \centering
        \includegraphics[width=\textwidth]{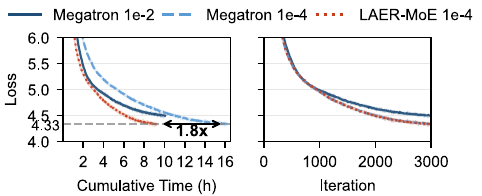}
        \caption{\small{Loss curves over training time (left side) and over training steps (right side) with different auxiliary loss weights.}}
    \end{subfigure}
    \vspace{0.5em}
    \begin{subfigure}[b]{0.47\textwidth}
        \centering
        \includegraphics[width=\textwidth]{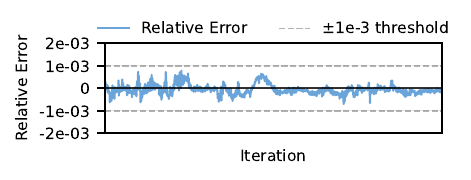}
        \caption{\small{Relative error between \name and Megatron with an auxiliary loss weight of 1e-4.}}
    \end{subfigure}
    \caption{\small{Convergence study on Mixtral-8x7B e8k2.}
    }
    \label{fig:convergence_study}
\end{figure}

Last but not least, we conduct convergence experiments on Mixtral-8x7B e8k2 with 4K sequence length. We set \name's weight of auxiliary loss to 1e-4 and compare it with Megatron using 1e-2 and 1e-4 as the weight of auxiliary loss. 
As shown in Fig.~\ref{fig:convergence_study}(a), with an auxiliary loss of 1e-4, our system converges at the same rate as Megatron (relative error < 1e-3, see Fig.~\ref{fig:convergence_study}(b)), demonstrating its correctness. 
Coupled with the numerical precision guarantees detailed in Sec.~\ref{fsep}, this convergence behavior confirms that FSEP incurs no loss in training precision.
Besides, although training with auxiliary loss of 1e-4 requires fewer iterations than 1e-2, Megatron's 1e-2 auxiliary loss weight improves routing load balance, resulting in faster iterations than 1e-4, thus achieving faster convergence. However, \name can train rapidly with an auxiliary loss of 1e-4, achieving the best convergence speed.

\subsection{Case Study}
\label{sec:case_study}

\begin{figure}[t]
    \centering
    \includegraphics[width=0.47\textwidth]{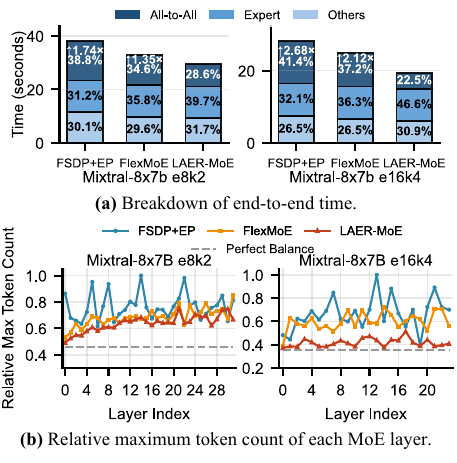}
    \caption{\small{Case study of breakdown and maximum token count.}}
    \label{fig:case_study}
\end{figure}

To clearly analyze the optimization sources of \name, we conduct an in-depth case study on Mixtral-8x7b for certain iterations using the Wikitext dataset. We break down the end-to-end time (averaged across all ranks), highlighting the All-to-All component, as shown in Fig.~\ref{fig:case_study}(a). 
The breakdown reveals that the speedup mainly comes from the acceleration of All-to-All communication, while expert computation and others (including attention computation and memory operations before and after All-to-All) have similar times across different methods.
For FSDP+EP, due to tail latency from load imbalance, All-to-All communication time reached up to 40\%. 
FlexMoE reduces this proportion through load balancing methods, while \name achieves a reduction of All-to-All communication to below 20\% by employing a more flexible load balance planner, resulting in up to a 2.68$\times$ speedup in All-to-All communication compared to the baseline.
This evidence validates that our approach effectively facilitates load balancing, thereby translating directly into enhanced training efficiency.
Regarding the breakdown details, the ``Others'' component primarily consists of attention layer operations and memory operations before and after the All-to-All phase. Notably, the ``Others'' part of these methods has a smaller proportion compared to Megatron (Fig.~\ref{fig:token_time_distribution}(b)), because Megatron employs tensor parallelism for attention layers to handle memory consumption due to insufficient FSDP support, resulting in a higher proportion of time in ``Others'' due to TP's communication overhead.

To further investigate the root cause, we analyze the maximum token count per device per layer across different methods, as illustrated in Fig.~\ref{fig:case_study}(b). 
The grey dashed line represents the token count for perfect balance. \name consistently achieves minimal deviation from the ideal balance across all scenarios.
In contrast, FlexMoE, which continuously adjusts previous expert layouts, may suffer from suboptimal adjustments when load changes, leading to suboptimal load balancing on e16k4. Meanwhile, \name optimizes globally at each iteration, and with the e16k4 models increasing the number of experts per device, it provides more flexible optimization opportunities, achieving nearly perfect load balancing.

\subsection{Planner Performance}
\label{sec:planner_performance}

\begin{table}[t]
    \centering
    \small
    \caption{Performance of lite routing.}
    \label{tab:lite_routing_performance}
    \begin{tabular}{ccc}
        \toprule
        Model & Lite Routing & Percentage of \\
         & Time(ms) & Total Time \\
        \midrule
        Mixtral-8x7B e8k2 & 24.965 & 0.084\% \\
        Mixtral-8x7B e16k4 & 30.994 & 0.094\% \\
        \bottomrule
    \end{tabular}
\end{table}

\begin{figure}[t]
    \centering
    \includegraphics[width=0.47\textwidth]{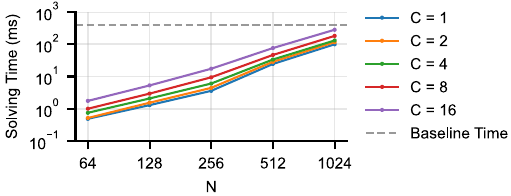}
    \caption{\small{Performance of expert layout solver.}}
    \label{fig:expert_layout_solver}
\end{figure}

We also analyze the efficiency of our planner algorithm.
For lite routing, we measure the time consumed by each iteration's related operations as a percentage of the total time, as shown in Tab.~\ref{tab:lite_routing_performance}. 
The percentage is less than 0.1\%, indicating that although the lite routing must execute synchronously during training, it has almost no additional impact on end-to-end performance.
The complexity of the replica allocation algorithm is $O(NC \log (NC))$, and the complexity of the expert placement algorithm is $O(N^2C)$, thus the complexity of the expert layout algorithm is $O(|\epsilon|N^2C)$. 
We fix $|\epsilon|$ to 2 (proportional and even allocation) and scale $N$ and $C$ to obtain the solving times shown in Fig.~\ref{fig:expert_layout_solver}, where the grey dashed line represents the baseline time, i.e., which is the average total time consumed per transformer layer in Mixtral-8x7b-e8k2.
As the cluster scale increases, the time per transformer layer exceeds the baseline time, yet even when scaled to 1024 GPUs, our algorithm remains highly efficient, consistently below the baseline time. 
Moreover, for larger $C$ and $N$, since solving is layer-independent, we can parallelize solvers for different layers across multiple CPU processes to relax the real-time constraints.
Thus, the planning will never become a bottleneck.

\subsection{Ablation Study}
\label{sec:ablation_study}

\begin{figure}[t]
    \centering
    \includegraphics[width=0.47\textwidth]{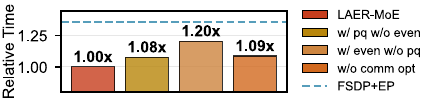}
    \caption{\small{Ablation study on Mixtral-8x7B e8k2.}}
    \label{fig:ablation_study}
\end{figure}

To validate our expert layout solver and communication optimizations in \name, we conduct ablation studies Mixtral-8x7b e8k2, as shown in Fig.~\ref{fig:ablation_study}. 
We compare \name with incomplete solving algorithms that employ only a single expert replication scheme, where 'pq' denotes the priority queue proportional allocation scheme and 'even' indicates the even allocation scheme.
It is evident that relying solely on one scheme cannot effectively handle all routing distribution scenarios, while utilizing multiple replication schemes offers robust performance. 
For communication optimization, we compare \name with a system lacking the communication optimizations in Fig.~\ref{fig:fsep_communication}, revealing that without such optimizations, the overlap effect of communication is compromised, subsequently affecting end-to-end efficiency.
\section{Related Work}

\textbf{Load Balancing in MoE Training.} 
FasterMoE~\cite{fastermoe} broadcasts hot experts to all devices, introducing extra expert communication.
SmartMoE~\cite{smartmoe} reassigns experts to co-locate hot and cold ones on the same device to alleviate load imbalance, but only changes expert locations without allowing replication. 
Prophet selectively replicates hot experts across nodes, yielding skewed parameter traffic proportional to replica count. 
FlexMoE~\cite{flexmoe} combines replacement and replication, dynamically tuning replica counts and locations of experts, but still cannot resolve the additional communication of expert parameters.
Our newly proposed parallel paradigm is completely orthogonal to these methods and offers more flexible expert layout choices.
Crucially, \name supports per-iteration re-layout, enabling real-time adaptation even in scenarios with rapidly shifting expert hotspots. This capability ensures sustained training efficiency, offering a decisive advantage over prior approaches that are limited to infrequent, coarse-grained adjustments.

\textbf{Communication and Computation optimizations.} 
MegaBlocks~\cite{megablocks} models expert GEMMs as sparse matrix multiplication and develops custom kernels. 
DeepSpeedMoE and MoESys~\cite{deepspeed-moe,moesys} reduce communication pairs via hierarchical optimizations.
NetMoE~\cite{netmoe} reorders sample placement to reduce inter-node communication. 
DeepEP~\cite{deepep} provides specialized All-to-All kernels. 
FasterMoE~\cite{fastermoe}, Tutel~\cite{tutel}, ScheMoE~\cite{schemoe}, MPMoE~\cite{mpmoe}, and PipeMoE~\cite{pipemoe} chunk All-to-All and expert compute to overlap communication and computation. 
Comet~\cite{comet} further fuses these into a single GPU kernel.
HiDup~\cite{hidup}, Lancet~\cite{lancet}, Lina~\cite{lina}, DeepSeek-V3~\cite{deepseek-v3} refine the scheduling of communication and computation to achieve better overlap. 
All are orthogonal to our approach and can be composed with our work for higher throughput.

\section{Discussion}

\textbf{Performance in Balanced Scenarios.}
As detailed in Sec.~\ref{fsep}, our method's communication volume is nearly identical to FSDP+EP. Consequently, in balanced expert load scenarios, the performance of \name is comparable to FSDP+EP. However, we emphasize that \name is architected primarily to address imbalanced scenarios. Our core objective is to decouple system efficiency from algorithmic constraints, providing researchers with the flexibility to explore algorithms with lower auxiliary loss weights (which potentially yield better model quality) while maintaining high training throughput and system robustness.

\textbf{Global Peak Memory Analysis.}
In Sec.~\ref{sec:case_study}, we report the relative maximum token count for different methods across various layers. This metric serves as an indirect proxy for per-layer activation memory usage (since memory consumption of activations is linear w.r.t. token counts). Since the heavy experts often differ from one layer to the next, FSDP+EP exhibits relatively balanced global peak memory across different ranks. Consequently, our method's impact on the global peak memory is not significant.
This does not conflict with our optimization goal: our method is designed to balance the computational load within each layer.

\textbf{Scalability on Large-scale Clusters.}
While empirical experiments on large-scale clusters are limited by resource constraints, we provide a theoretical analysis to demonstrate the scalability of \name. 
First, as cluster size increases, the reduced per-device memory footprint allows for larger micro-batch sizes, effectively compensating for a potential decrease in bandwidth ($V_{comm}$) and maintaining computation-communication overlap (Eq.~\ref{eq:computation-communication-overlap}). 
Second, in cross-rack scenarios where bandwidth is typically constrained, \name is compatible with hybrid parallelism (e.g., Pipeline Parallelism), which can mitigate limited cross-rack bandwidth by confining All-to-All communication within racks to preserve training efficiency. 
Finally, we conducted simulations using real routing traces from Mixtral-8x7B-e8k2 training (detailed in Appendix~\ref{sec:scalability_analysis}), which indicate that the speedup yielded by our re-layout algorithm remains stable as the cluster scales from 8 to 128 GPUs, demonstrating that our approach scales efficiently with cluster expansion.

\section{Conclusion}

We presented \name, a novel system for distributed MoE training that addresses load balancing challenges.
We proposed Fully Sharded Expert Parallelism (FSEP) paradigm, which enables flexible, in-training expert re-layout while removing re-layout communication overhead.
We also designed an intelligent load balancing planner that dynamically co-optimizes token routing and expert layout, improving throughput without sacrificing model quality.
Experiments show up to $1.69\times$ acceleration over SOTA distributed systems, advancing efficient large-scale MoE training.

\section*{Acknowledgments}
This work is supported by National Natural Science Foundation of China (U23B2048, 62402011), Fundamental and Interdisciplinary Disciplines Breakthrough Plan of the Ministry of Education of China (JYB2025XDXM108), ByteDance-PKU joint program, and High-performance Computing Platform of Peking University. Fangcheng Fu and Bin Cui are the corresponding authors.


\bibliographystyle{ACM-Reference-Format}
\bibliography{acmart}

\appendix
\section{Artifact Appendix}

\subsection{Abstract}

This artifact includes codes and scripts for reproducing all experiments in the paper. The artifact contains the implementation of LAER-MoE along with baseline implementations (Megatron-LM, FSDP-EP, FLEX-MoE), training scripts for various MoE models (Mixtral-8x7B, Mixtral-8x22B, Qwen-8x7B), and evaluation scripts to reproduce all experimental results including end-to-end performance (Figure 8), convergence study (Figure 9), case study (Figure 10), planner performance (Figure 11), and ablation study (Figure 12).

\subsection{Artifact check-list (meta-information)}

{\small
\begin{itemize}
  \item {\bf Program: } Python, PyTorch, CUDA
  \item {\bf Compilation: } NVCC 12.1, CMake >= 3.21
  \item {\bf Model: } Mixtral-8x7B, Mixtral-8x22B, Qwen-8x7B with different expert configurations (e8k2, e16k4)
  \item {\bf Data set: } WikiText-103, C4
  \item {\bf Run-time environment: } Python 3.9.2, CUDA 12.1, NCCL 2.18.1
  \item {\bf Hardware: } 4-node GPU cluster, each node with 8 NVIDIA A100 80GB GPUs, NVLink (300 GB/s intra-node), Infiniband (800 Gbps inter-node)
  \item {\bf Execution: } Distributed training using torch.distributed.launch
  \item {\bf Metrics: } End-to-end performance (Training throughput (tokens/s))
  \item {\bf Output: } End-to-end trainging throughput, loss curves, and figures.
  \item {\bf Experiments: } End-to-end, convergence study, case study, planner performance, ablation study
  \item {\bf How much disk space required (approximately)?: } ~200 GB for per model checkpoints, 2TB in total.
  \item {\bf How much time is needed to prepare workflow (approximately)?: } 2-4 hours for environment setup and installation.
  \item {\bf How much time is needed to complete experiments (approximately)?: } 0.5-1 hour per-configuration for most experiments, ~10 hours per-configuration for each convergence study.
  \item {\bf Publicly available?: } Yes (Anonymized link provided for review)
  \item {\bf Code licenses (if publicly available)?: } Apache 2.0
  \item {\bf Workflow automation framework used?: } Bash scripts
  \item {\bf Archived (provide DOI)?: } 10.5281/zenodo.18298795
\end{itemize}
}

\subsection{Description}

\subsubsection{How to access}

The artifact is available on GitHub repository(\url{https://github.com/Fizzmy/LAER-MoE-AE}) and Zenodo (\url{https://doi.org/10.5281/zenodo.18298795}), including installation instructions, experiment workflow, and expected results.

\subsubsection{Hardware dependencies}

We conduct our experiments on a 4-node GPU cluster, with each node containing 8 NVIDIA A100 80GB GPUs. Within nodes, GPUs are connected via NVLink with peak unidirectional communication bandwidth of 300 GB/s intra-node. Nodes are interconnected via Infiniband with 800 Gbps bandwidth inter-node.

\subsubsection{Software dependencies}

The following software dependencies are required:
\begin{itemize}
  \item Python 3.9.2
  \item CUDA 12.1
  \item PyTorch 2.1.0 (cu121)
  \item Flash-Attention v2.5.8
  \item Apex (commit 312acb4)
  \item Transformer-Engine (commit 7f2afaa, for baseline only)
  \item CMake >= 3.21 (for baseline only)
  \item Additional Python packages: packaging, ninja, bsutil, pybind11, einops
\end{itemize}

\subsubsection{Data sets}

We use two datasets for evaluation:
\begin{itemize}
  \item WikiText-103: A language modeling dataset containing over 100 million tokens from Wikipedia articles
  \item C4: A colossal, cleaned version of Common Crawl's web crawl corpus
\end{itemize}
The datasets are preprocessed using the Mixtral tokenizer.

\subsubsection{Models}

We evaluate on the following MoE model configurations:
\begin{itemize}
  \item Mixtral-8x7B: e8k2 and e16k4 configurations
  \item Mixtral-8x22B: e8k2 and e16k4 configurations
  \item Qwen-8x7B: e8k2 and e16k4 configurations
\end{itemize}
where e8k2 means 8 experts with top-2 routing, and e16k4 means 16 experts with top-4 routing.

\subsection{Installation}

Users need to create a virtual environment and install PyTorch, Flash-Attention, Apex, and Transformer-Engine. Please refer to \texttt{README.md} for detailed instructions. To install \name, use the following commands:

\begin{lstlisting}[style=bashstyle, language=bash]
cd LAER-MoE
pip install -r requirements.txt
pip install -e . --no-build-isolation
\end{lstlisting}

\subsection{Experiment workflow}

All experiments should be conducted from the \texttt{./LAER-MoE} directory.

\subsubsection{Prepare Datasets and Checkpoints}

For datasets preparation, users should preprocess the datasets using Mixtral tokenizer. Please refer to \texttt{README.md} for detailed preprocessing instructions.

For checkpoints preparation, users should generate checkpoints for each model configuration before running experiments. Please refer to \texttt{\name/README.md} for detailed instructions on checkpoint generation.

\subsubsection{Performing Experiments}

We provide shell scripts to perform experiments for each figure in the paper. All scripts are located in \texttt{LAER-MoE/scripts\_ae/}.

\textbf{End-to-End Performance (Figure 8):}
\texttt{e2e.sh} will start the end-to-end evaluation in \S\ref{sec:end-to-end}. Users can run the following command for each configuration:
\begin{lstlisting}[style=bashstyle, language=bash]
bash scripts_ae/e2e.sh <approach> <model_name> <aux_loss> <dataset>
\end{lstlisting}
where:
\begin{itemize}
  \item \texttt{<approach>}: \texttt{LAER}, \texttt{FLEX}, \texttt{FSDP}, or \texttt{megatron}
  \item \texttt{<model\_name>}: one of
  \begin{itemize}[nosep,leftmargin=3em]
    \item \texttt{mixtral-8x7b-e8k2}
    \item \texttt{mixtral-8x7b-e16k4}
    \item \texttt{mixtral-8x22b-e8k2}
    \item \texttt{mixtral-8x22b-e16k4}
    \item \texttt{qwen-8x7b-e8k2}
    \item \texttt{qwen-8x7b-e16k4}
  \end{itemize}
  \item \texttt{<aux\_loss>}: \texttt{0.0} or \texttt{1e-4}
  \item \texttt{<dataset>}: \texttt{wikitext} or \texttt{C4}
\end{itemize}

We also provide a script to run the end-to-end evaluation for all the models and all the approaches. Users can run the following command:
\begin{lstlisting}[style=bashstyle, language=bash]
bash scripts_ae/e2e_all.sh
\end{lstlisting}

Users can modify the \texttt{e2e\_all.sh} script to run the end-to-end evaluation for different models and different approaches.

\textbf{Convergence Study (Figure 9):}
\texttt{convergence.sh} will start the convergence study in \S\ref{sec:end-to-end}. Users can run the following command:
\begin{lstlisting}[style=bashstyle, language=bash]
bash scripts_ae/convergence.sh <approach> <aux_loss> <iter>
\end{lstlisting}
where \texttt{(<approach>, <aux\_loss>)} can be \texttt{(LAER, 1e-4)}, \texttt{(Megatron, 1e-2)}, or \texttt{(Megatron, 1e-4)}. Set \texttt{<iter>} to 3000 for full results or 1500 for faster artifact evaluation.

\textbf{Case Study (Figure 10):}
We provide two scripts to perform case study for breakdown and maximum token counts.

\textbf{\textit{Breakdown (Figure 10a):}}
\texttt{breakdown.sh} will start the breakdown study in \S\ref{sec:case_study}. Users can run the following command:
\begin{lstlisting}[style=bashstyle, language=bash]
bash scripts_ae/breakdown.sh <approach> <model_name>
\end{lstlisting}

\textbf{\textit{Maximum token counts (Figure 10b):}}
\texttt{token\_counts.sh} will start the maximum token counts study in \S\ref{sec:case_study}. Users can run the following command:
\begin{lstlisting}[style=bashstyle, language=bash]
bash scripts_ae/token_counts.sh <approach> <model_name>
\end{lstlisting}

where \texttt{<approach>} is \texttt{LAER}, \texttt{FLEX}, or \texttt{FSDP}, and \texttt{<model\_name>} is \texttt{mixtral-8x7b-e8k2} or \texttt{mixtral-8x7b-e16k4}.

\textbf{Planner Performance (Figure 11):}
\texttt{planner.sh} will start the planner performance study in \S\ref{sec:planner_performance}. Users can run the following command:
\begin{lstlisting}[style=bashstyle, language=bash]
bash scripts_ae/planner.sh <N> <C>
\end{lstlisting}
where \texttt{<N>} is the number of devices and \texttt{<C>} is the capacity of experts per device. To get average time for all (N, C) pairs, run without arguments:
\begin{lstlisting}[style=bashstyle, language=bash]
bash scripts_ae/planner.sh
\end{lstlisting}

\textbf{Ablation Study (Figure 12):}
\texttt{ablation.sh} will start the ablation study in \S\ref{sec:ablation_study}. Users can run the following command:
\begin{lstlisting}[style=bashstyle, language=bash]
bash scripts_ae/ablation.sh <approach>
\end{lstlisting}
where \texttt{<approach>} is \texttt{LAER}, \texttt{no\_even}, \texttt{no\_pq}, \texttt{no\_comm\_opt}, or \texttt{FSDP}.

\subsubsection{Plotting Figures}
To generate figures from experimental results:
\begin{lstlisting}[style=bashstyle, language=bash]
bash scripts_ae/plot.sh <figure_id> <type>
\end{lstlisting}
where \texttt{<figure\_id>} is \texttt{8}, \texttt{9}, \texttt{10a}, \texttt{10b}, \texttt{11}, or \texttt{12}, and \texttt{<type>} is \texttt{default} (use paper data) or \texttt{new} (use newly generated experimental results).

\subsection{Evaluation and expected results}

The absolute performance numbers may vary across different hardware and network configurations. However, we expect that the relative performance trends shown in Figures 8-12 should be reproducible, with the average speedup ratios remaining comparable to those reported in the paper.

\subsection{Experiment customization}

Users can customize the evaluation scripts to test the system performance on other workloads with different model architectures, datasets, or training parameters. Please refer to \texttt{\name/README.md} for detailed instructions on customization options and parameters.

\subsection{Notes}

\begin{itemize}
  \item The experiments require a 4-node cluster with 32 A100 GPUs. Results may vary on different hardware configurations and bandwidth.
  
  \item To ensure comparable training convergence across different approaches (with relative loss error below 1e-3), we save checkpoints for each model configuration. This requires significant disk space (approximately 200 GB per model). Users can choose to train without loading checkpoints, but this may result in different routing results and loss trajectories.
\end{itemize}

\balance

\section{Lite Routing Algorithm}
\label{sec:append_routing}
\begin{algorithm}
\caption{\small{Lite Routing Algorithm}}
\small
\label{algo:lite_routing}
\LinesNumbered
\KwIn{$R_{rank,:}$, $A_{rank,:}$, $N$, $E$, $C$, $rank$: rank of current device}
\KwOut{$S_{rank,:,:}$}
\For{each expert $j \in [0, E-1]$}{
    $rep_{intra} \leftarrow \text{set of replicas} j \text{ intra node}$\;
    $rep_{inter} \leftarrow \text{set of replicas} j \text{ inter node}$\;
    \If{$len(rep_{intra}) > 0$}{
        \For{$k \in rep_{intra}$}{
            $S_{rank,j,dev_k} \leftarrow S_{rank,j,dev_k} + \frac{R_{rank,j}}{len(rep_{intra})}$\;
        }
    }
    \Else{
        \For{$k \in rep_{inter}$}{
            $S_{rank,j,dev_k} \leftarrow S_{rank,j,dev_k} + \frac{R_{rank,j}}{len(rep_{inter})}$\;          
        }
    }
}
\Return $S_{rank,:,:}$\;
\end{algorithm}

Here, we illustrate the lite routing algorithm used in the token dispatcher. As shown in Alg.~\ref{algo:lite_routing}, this algorithm is performed separately in each device, and for each expert, if replicas are present within a node, tokens routed to the expert are evenly distributed among all intra-node replicas (Line~5-6). In the absence of intra-node replicas, tokens are evenly allocated among all replicas across global nodes (Line~8-9).

\section{Expert Replica Allocation Algorithm}
\label{sec:append_replica_alloc}

The replica allocation algorithm is illustrated in Alg.~\ref{algo:replica_allocation}. Specifically, the algorithm maintains a priority queue (Line~2-4) that identifies the expert with the highest average load (i.e., the ratio of actual load to the number of replicas) and increases the number of replicas for that expert until the total number of replicas meets the required threshold (Line~5-8).

\begin{algorithm}
\caption{\small{Replica Allocation Algorithm}}
\label{algo:replica_allocation}
\small
\LinesNumbered
\KwIn{$R$, $N$, $E$, $C$}
\KwOut{$expert\_rep$}
$expert\_rep, expert\_load \leftarrow \text{ones}(E), R.sum(axis=0)$\;
$priority\_queue \leftarrow \emptyset$\;
\For{$i \in [0, E-1]$}{
    $priority\_queue.push(i, \frac{expert\_load[i]}{expert\_rep[i]})$\;
}
\While{$sum(expert\_rep) < N \times C$}{
    $e \leftarrow \text{pop from } priority\_queue$\;
    $expert\_rep[e] \leftarrow expert\_rep[e] + 1$\;
    $priority\_queue.push(e, \frac{expert\_load[e]}{expert\_rep[e]})$\;
}
\Return $expert\_rep$\;
\end{algorithm}

\section{Scalability Analysis}
\label{sec:scalability_analysis}
To evaluate the scalability of \name beyond the physical constraints of our available hardware, we conducted a trace-driven simulation. 
This simulation utilizes real routing traces collected from the training of the Mixtral-8x7B-e8k2 model. 
We modeled the execution latency of the MLP module (including expert computation and All-to-All communication) under varying cluster sizes, ranging from 8 to 128 GPUs. 
Tab.~\ref{tab:scalability_sim} presents the simulation results. 
We observe that the speedup of the MLP module remains highly stable as the cluster size increases. 
This stability confirms that \name scales efficiently to large-scale clusters.
\begin{table}[h]
    \centering
    \caption{Simulated MLP Speedup of \name on varying cluster sizes using Mixtral-8x7B-e8k2 routing traces.}
    \label{tab:scalability_sim}
    \begin{tabular}{c c}
    \toprule
    \textbf{Number of GPUs} & \textbf{MLP Speedup} \\
    \midrule
    8   & 1.491$\times$ \\
    16  & 1.490$\times$ \\
    32  & 1.488$\times$ \\
    64  & 1.487$\times$ \\
    128 & 1.482$\times$ \\
    \bottomrule
    \end{tabular}
\end{table}


\end{document}